\pdfoutput=1

\documentclass[english, sigconf, nonacm]{acmart}
\usepackage[utf8]{inputenc}
\usepackage[T1]{fontenc}

\makeatletter
    \def\balanceissued{unbalanced}
    \let\oldbibitem\bibitem
    \def\bibitem{%
        \ifnum\thepage=\getrefnumber{TotPages}%
            \expandafter\ifx\expandafter\relax\balanceissued\relax\else%
                \balance%
                \gdef\balanceissued{\relax}\fi%
            \else\fi%
        \oldbibitem}
\makeatother

\newcommand{\eg}{\emph{e.g.}}
\newcommand{\ie}{\emph{i.e.}}

\usepackage{quoting}
\usepackage{tikz}
\usepackage{tikzpeople}
\usetikzlibrary{backgrounds}
\usetikzlibrary{fit}
\usetikzlibrary{positioning}
\usetikzlibrary{shadows}

\title[Online publication of court records]{Online publication of court records: circumventing the privacy-transparency trade-off}

\author{Tristan Allard}
\affiliation{Univ Rennes, CNRS, IRISA}
\email{tristan.allard@irisa.fr}
\author{Louis B{\'e}ziaud}
\orcid{0000-0002-4974-3492}
\affiliation{Univ Rennes, CNRS, IRISA}
\affiliation{Université du Québec à Montréal}
\email{louis.beziaud@irisa.fr}
\author{S{\'e}bastien Gambs}
\affiliation{Université du Québec à Montréal}
\email{gambs.sebastien@uqam.ca}

\begin{abstract}
  The open data movement is leading to the massive publishing of court
  records online, increasing transparency and accessibility of
  justice, and to the design of legal technologies building on the
  wealth of legal data available. However, the sensitive nature of
  legal decisions also raises important privacy issues. Current
  practices solve the resulting privacy \emph{versus} transparency
  trade-off by combining access control with (manual or semi-manual)
  text redaction. 
  In this work, we claim that current practices are insufficient for coping with
  massive access to legal data (restrictive access control policies is
  detrimental to openness and to utility while text redaction is
  unable to provide sound privacy protection) and advocate for a
  integrative approach that could benefit from the latest developments
  of the privacy-preserving data publishing domain. We present a
  thorough analysis of the problem and of the current approaches, and
  propose a straw man multimodal architecture paving the way to a
  full-fledged privacy-preserving legal data publishing system.
\end{abstract}

\begin{document}

\maketitle

\section{Introduction}

The opening of legal decisions to the public is one of the cornerstones of many modern democracies: it allows to audit and make accountable the legal system by ensuring that justice is rendered with respect to the laws in place.
As stated in \cite{bentham1843works}, it can even be considered that \emph{``publicity is the very soul of justice''}.
Additionally, in countries following the common law, the access to legal decisions is a necessity as the law in place emerged from the previous decisions of justice courts.

Thus, it is not surprising that the transparency of justice is enshrined in many countries as a fundamental principle, such as the \emph{right to a public hearing} provided by the Article 6 of the European Convention on Human Rights, the Section~135(1) of the Courts of Justice Act (Ontario) stating the general principle that \emph{``all court hearings shall be open to the public''} or in Vancouver Sun (Re) \emph{``The open court principle has long been recognized as a cornerstone of the common law''}.
The open data movement push for free access to law with for example the Declaration on Free Access to Law~\cite{fal}.
Multiple open government initiatives also consider the need for an open justice, such as the ``Loi pour une République numérique'' in France, the Open Government Partnership, the Open Data Charter, the Canada's Action Plan on Open Government.
This trend is studied in a report of the OCDE~\cite{organisation_for_economic_co-operation_and_development_call_2011}, in~\cite{mcdermott2010building} for the USA or~\cite{mcclean2011not} for the UK.

Combined with recent advances in machine learning and natural language processing, the (massive) opening of legal data allows for new practices and applications (called legal technologies).
Nonetheless, not all legal decisions should directly be published as such due to the privacy risks that might be incurred by victims, witnesses, members of the jury and judges.
Some privacy risks have been considered and mitigated by legal systems for a long time.
For instance, the identities of the individuals involved in sensitive cases, such as cases with minors, are usually \emph{anonymized} by default because they belong to a vulnerable subgroup of the population.
In situations in which the risks of reprisal are high (\eg, terrorism or organized crimes cases), judges, lawyers and witnesses might also ask for their identities to be hidden~\cite{jacquin2017peur, fleuriot2017avec}.
Finally, the identities of the members of a jury are also usually protected to guarantee that they will not be coerced but also to ensure that the strategy deployed by the lawyers is not tailored based on their background.
Legal scholars are aware of the need for privacy when opening sensitive legal reports~\cite{conleySustainingPrivacyOpen, jaconelli2002open, bailey2016revisiting}.

In the past, these privacy risks were limited due to the efforts that were required to access the decisions themselves.
For instance, some countries require to go directly to the court itself to be able to access the legal decisions.
Even when the information is available online, the access to legal decisions is usually on a one-to-one basis through a public but restricted API rather than enabling a direct download of the whole legal corpus.
Typical restriction mechanisms include CAPTCHAs (SOQUIJ), quotas (CanLII), registration requirement as well as policy agreement and limitation of access to research scholars (Caselaw).
Furthermore, the fact that a legal decision is public does not mean that it can, legally, be copied and integrated in other systems or services without any restrictions.

A first approach to limit the privacy risks consists in \emph{redacting} the legal decisions before publishing them.
Redaction mostly follows predefined rules that list the information that must be removed or generalized and define how (\eg, by replacing the first and last names by initials, by a pseudonym)~\cite{opijnen2017line}.
Redaction is in general semi-manual (and sometimes fully manual) because automatic redaction is error-prone~\cite{marrero2013named}.
This makes it extremely costly, not scalable, and does not completely remove the risks of errors~\cite{opijnen2017line}.
For example, 3.9 million decisions are pronounced in France every year but only 180000 are recorded in government databases and less than 15000 are made accessible to the public~\cite{etalab-openjustice}.
Moreover, even a perfect redaction would still offer weak privacy guarantees.
A redacted text still contains a non-negligible amount of information, possibly identifying or sensitive, that may be extracted, \eg, from the background of the case or even from the natural language semantics.

Another approach is access control, such as non-publication (\eg, a case involving terrorism was held in secret in Britain~\cite{secret-trial-britain}), rate limit, or registration requirements. However, access control mechanisms are binary and do not protect the privacy of the texts for which the access is authorized.
Furthermore, restricting massive accesses for blocking also restricts the development of legal technologies that require a massive access to legal data.

In a nutshell, this paper makes the following contributions:
\begin{itemize}
    \item We clearly state the problem of reconciling transparency with privacy when opening legal data massively.
    \item We analyze the limits of the current approach, widespread in real-life.
    \item We propose a high-level straw man architecture of a system for publishing legal data massively in a privacy-preserving manner without precluding the traditional open court principles.
\end{itemize}

The outline of the paper is as follows.
First in Section~\ref{sec:problem}, we state the problem by describing precisely the content of legal data and by explaining the open legal data desiderata.
Afterwards, we present the privacy limitations of the current approach, redaction, in Section~\ref{sec:limits}, before describing in Section~\ref{sec:contrib} our proposal of architecture for the publication of legal data ensuring both privacy and utility.
Finally, we conclude in Section~\ref{sec:conclusion}.

\section{Problem statement}\label{sec:problem}

\subsection{Legal data}

Legal reports are defined as written documents produced by a court about a particular judgment, which is itself a written decision of a court regarding a particular case (oral judgments are transcribed).
Although the content of a case report varies with respect to the court and the country, it can consist of elements such as~\cite{brief-case}:
\begin{enumerate}
    \item the case name and case citation (identifier);
    \item the date of judgment and the hearing dates;
    \item the court and judges involved in the decision;
    \item the appearances (parties and their representatives);
    \item the statement of facts: identify--sometimes in great length--the relationship and status of the parties, the legally relevant facts (\ie, what happened), and the procedurally significant facts (\eg, cause of action, relief request, raised defenses);
    \item the procedural history: describes--if applicable--the disposition of the case in the lower court(s), the damages awarded, the reason for appeal, etc.;
    \item the issues: point of law in dispute;
    \item the law of the case: elements of law that the court applies;
    \item the concurring and/or dissenting opinions (of judges);
    \item the orders: the decision itself.
\end{enumerate}

We can broadly distinguish three different categories of judicial data depending: metadata, facts and reasoning.
Metadata (elements 1, 2, 3 and 4) correspond to identifiers of the case and basic information (\eg, date, parties and judge) and is written mostly in a structured way.
Facts (elements 5, 6 and 7) are information pertaining to the parties, disclosing their personal ``story''.
Reasoning (elements 8, 9 and 10) is the logic of the case, which is not specific to the parties.

\subsection{Desiderata for the opening of legal data}

\subsubsection{Need for readability and accessibility}\label{sec:readability}

The access to legal decisions is required both for ethical (transparency) and practical reasons such as case law, which is the use of past legal decisions to support the decision for future cases.
Thus, the judiciary system is built on the assumption that legal decisions are made public and accessible by default (\emph{open-court principle}), so that (1) citizens are able to inspect decisions as a way to audit the legal system and (2) past decisions can be used to interpret laws, and as such must be known from legal practitioners and citizens.
It follows that decisions must be made available in a form readable by humans (\ie, natural language).
Natural language format can be opposed to machine-readable formats such as word-vectors representation or logical propositions, which we will discuss later.
The need for openness, the current practice in terms of open court, and the associated risks are detailed in~\cite{conleySustainingPrivacyOpen, martin2008online}.
They conclude that, although there are powerful voices in favor of open court, radical changes in access and dissemination require new privacy constraints, and a public debate on the effect of sharing and using information in records.

Accessibility is also an important issue.
In the past, the access to decisions required attending public hearings or reading books called ``reporters''.
Later, decisions have started to be shared on digital medium such as compact discs or DVDs for example before being accessible online more recently.
For instance, in the USA, CourtListener\footnote{\url{https://www.courtlistener.com}} shares 3.6M decisions and the Caselaw access project\footnote{\url{https://case.law}} 6.7M unique cases; the Canadian Legal Information Institute\footnote{\url{https://www.canlii.org}} (CanLII) publishes 2.5M Canadian decisions.
The aim of these services is to facilitate access to legal records to individuals--law professionals (judges, lawmakers and lawyers), journalists, or citizens.
The online publication also enables the large-scale access and processing of records, in particular due to the standardized format.

\subsubsection{Need for massive accesses (legal technologies)}\label{sec:legaltech}

The term \emph{legal technologies} broadly encompasses all the technologies used in the context of justice.
The website CodeX Techindex\footnote{\url{http://techindex.law.stanford.edu}}, a project by the Stanford Center for Legal Informatics, references more than a thousand companies, and defines nine different categories: (1) Marketplace, (2) Document Automation, (3) Practice Management, (4) Legal Research, (5) Legal Education, (6) Online Dispute Resolution, (7) E-Discovery, (8) Analytics and (9) Compliance.

A subset of these categories--2, 4, 7, 8 and 9--requires some form of ``understanding''  of legal documents, usually performed through natural language processing (NLP) and machine learning (ML) approaches~\cite{praduroux2016legal,dale_law_2019}.
We focus here on these categories as they are based on the analysis of a large number of legal data.
One of the main challenges we have faced is that usually companies provide very few technical details about their actual processing and usage of legal documents.

The automatic processing and analysis of legal records have multiple applications, such as computing similarity between cases~\cite{thenmozhi2017text, MINOCHA189, 10.1145/3140107.3140119}, predicting legal outcomes~\cite{aletras2016predicting, katz2017general} (\eg, by weighing the strength of the defender arguments and the legal position of a client in a hypothetical or actual lawsuit), identifying influential cases~\cite{marquesMachineLearningExplaining2019, movzina2005argument, siegel2017cara} or important part of laws~\cite{mokanov2019facts2law}, estimating the risk of recidivism~\cite{tan2018investigating}, summarizing legal documents~\cite{10.1007/978-3-642-13059-5_8}, extracting entities (\eg, parties, lawyers, law firms, judges, motions, orders, motion type, filer, order type, decision type and judge names) from legal documents~\cite{Quaresma2010, custisWestlawEdgeAI2019},  topic modelling~\cite{nallapati2008legal, 10.1007/s10506-009-9077-9}, concept mapping~\cite{10.1007/3-540-63233-6_501} or inferring patterns~\cite{kort1965quantitative, ashley2013toward}.

\paragraph{Focus on text-based legal techs}
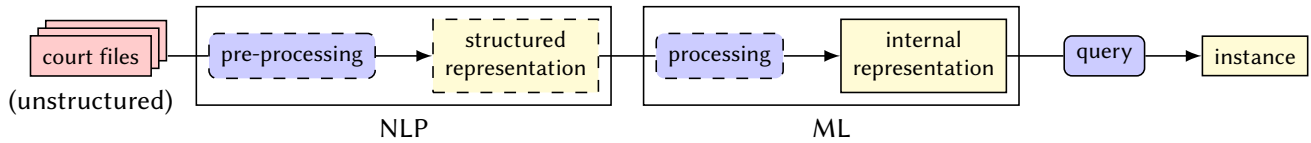
\begin{figure*}[htbp]
    \centering
    \resizebox{\linewidth}{!}{
        \begin{tikzpicture}[
    font=\sffamily\scriptsize,
    node distance=0.5cm,
    proc/.style={draw, fill=blue!20, rounded corners=2pt},
    obj/.style={draw, fill=yellow!20},
    tmp/.style={dashed},
    ]
    \node[fill=red!20, draw, double copy shadow] (files) {court files};
    \node[above=0 of files, minimum height=4pt, inner sep=0] {};
    
    \node[proc, tmp, right=of files] (preproc) {pre-processing};
    \node[obj, tmp, right=of preproc, align=center] (struct) {structured\\ representation};
    
    \node[proc, tmp, right=of struct] (proc) {processing};
    \node[obj, right=of proc, align=center] (irepr) {internal\\ representation};

    \node[proc, right=of irepr] (query) {query};
    \node[obj, right=of query, align=center] (instance) {instance};

    \node[fit=(preproc)(struct), draw, label=south:{\footnotesize NLP}] (nlp) {};
    \node[fit=(proc)(irepr), draw, label=south:{\footnotesize ML}] (ml) {};

    \draw[-latex] ([xshift=4pt]files.east) -- (preproc) -- (struct);
    \draw[-latex] (struct) -- (proc) -- (irepr);
    \draw[-latex] (irepr) -- (query) -- (instance);

    \node[below=0 of files, align=right] {\footnotesize (unstructured)};
\end{tikzpicture}
    }
    \caption{High-level pipeline of court files processing for Legal Techs}\label{fig:pipe}
\end{figure*}

Most of the technologies introduced in the previous section rely on the processing of large database of legal data.
However, the unstructured nature of legal data is one of the main challenges of the application of artificial intelligence in law~\cite{alarie2018artificial}.
Consequently, the analysis of a legal text corpus first requires to apply some pre-processing to add structure to the text.
Figure~\ref{fig:pipe} represents an abstract processing pipeline for court files, extracted mostly from academic papers\footnote{The majority of the legal technologies market consists in commercial applications. They do not give information about their inner working and underlying techniques.}, and inferred from the current practice of text analysis and descriptions of associated technologies.
In the following, we assume that any application involving the use of machine learning (as highlighted by most legal tech companies) is applied to court records.
The first NLP step transforms the unstructured data (\ie, natural language) into some structured representation (see below) by pre-processing it.
Afterwards, the second ML step corresponds to the actual application, which is the training (\ie, processing) of the ML algorithm, whose output is represented by the "internal representation" block. The term instance represents the output of the model given some query (\eg, applicable laws given a set of keywords representing infractions).

The pre-processing can be diverse and depends on the task (\eg, extracting a citation graph between cases).
However, most NLP-based applications usually rely on a text model.
Many models are statistical-based ones, such as document-word-frequency matrix, in which the corpus is decomposed into a matrix in which each cell contains the number of times a particular word appears in a document.
This model has multiple variations such as bag-of-words (BoW)~\cite{joachims1998text}, term frequency-inverse document frequency, or $n$-grams~\cite{zhang2015character}.
For example, a combination of those techniques are used in~\cite{aletras2016predicting} to predict decisions from the European Court of Human Rights, and by~\cite{kimStatuteLawInformation2019} to identify law articles given a query or to answer to questions given a law article.
More recent approaches follow a neural network architecture in which a model is trained on the corpus with the objective to predict a word given a context, which is called word embeddings~\cite{mikolov2013distributed}. Multiple variations of this structure exist~\cite{joulin2016bag, kim2014convolutional, liu2016recurrent, lai2015recurrent, zheng2018hierarchical}.
This approach has been used for example in \cite{marquesMachineLearningExplaining2019} to rank and explain influential aspects of law, or by~\cite{mokanov2019facts2law} to predict the most relevant sources of law for any given piece of text using ``neural networks and deep learning algorithms''.

\subsubsection{Need for privacy}\label{sec:privacy}

The massive opening of legal decisions for transparency and
technological reasons must not hinder the fundamental rights such as the right to privacy as emphasized by current open justice laws.
In particular in this setting, the privacy of a least
three main actors must be protected: namely the individuals directly involved in decisions (\ie, the parties), the individuals cited by decisions (\eg, experts or witnesses), and the individuals administering the laws (\ie, magistrates).

However, the problem of publishing legal decisions in a privacy-preserving manner is a difficult one.
For instance, authorship attacks~\cite{abbasi2008writeprints} may lead to the re-identification of magistrates behind written decisions, or the presence of \emph{quasi-identifiers}\footnote{A quasi-identifier is a combination of attributes that are usually unique in the population, thus indirectly identifying an individual.
A typical example is the triple \texttt{(age, zip code, gender)}.} within the text decisions may lead to the re-identification of the individuals involved in or cited.
Famous real-life examples, such as the governor Weld's~\cite{10.1142/S0218488502001648} or Thelma Arnold's re-identification~\cite{aol-arrington-tc-06}, both based on the exploitation of quasi-identifiers, are early demonstrations of the failure of naive privacy-preserving data publishing schemes.
Thus despite the fact that legal decisions are written as unstructured text, structured information can be extracted from them, including the formal argument, the decision itself (\eg, ``guilty'' or ``innocent''), as well as arbitrary information about the individuals involved (\eg, gender, age and social relationships).

\emph{Pseudonymization} schemes simply consist in removing directly identifying data (\eg, social security number, first name and last name, address) and keeping unchanged the rest of the information (quasi-identifiers included).
These schemes provide a very weak protection level, as  acknowledged in privacy legislations (\eg, GDPR), which has led to the development of new approaches for sanitizing personal data in the last two decades (see for instance the survey in~\cite{ppdp-chen-ft-09}).
In this paper, we focus on privacy-preserving data publishing schemes providing formal privacy guarantees that hold against several publications (as required by any real-life privacy-preserving data publishing system).
These schemes are based on (1) \emph{a formal model} stating the privacy guarantees the scheme as well as \emph{a privacy parameter} for tuning the ``privacy level'' that must be achieved, and (2) \emph{a sanitization algorithm} designed to achieve the chosen model.

A formal model exhibits a set of \emph{composability properties} that defines formally the impact on the overall privacy guarantees of using the scheme on a \emph{log of publications} (also called \emph{disclosures log} in the following).
In particular, we will consider the $\epsilon$-differential privacy model~\cite{Dwork:2006:DP:2097282.2097284}, defined formally in Definition~\ref{def:dp}, parametrized by $\epsilon$, and achievable by the Laplace mechanism.
Its self-composability properties are stated in Theorem~\ref{the:composition} and its overall privacy guarantees are quantified by the evolution of the disclosures log, and in particular by the evolution of the $\epsilon$ value along the various differentially-private releases.

\begin{definition}[$\epsilon$-differential privacy \cite{Dwork:2006:DP:2097282.2097284}]\label{def:dp}
    The randomized function $\mathtt{f}$ satisfies $\epsilon$-differential privacy, in which $\epsilon > 0$, if:
    \[ \mathtt{Pr} [ \mathtt{f} ( \mathcal{D}_1 ) = \mathcal{O} ] \leq e^\epsilon \cdot \mathtt{Pr} [ \mathtt{f} ( \mathcal{D}_2 ) = \mathcal{O} ] \]
    for any set $\mathcal{O}\in Range(\mathtt{f})$ and any tabular dataset $\mathcal{D}_1$ and $\mathcal{D}_2$ that differs in at most one row (in which each row corresponds to a distinct individual).
\end{definition}

In a nutshell, $\epsilon$-differential privacy ensures that the presence (or absence) of data of a single individual has a limited impact on the output of the computation, thus limiting the inference that can be done by an adversary about a particular individual based on the observed output.

\begin{theorem}[Sequential and parallel Composability~\cite{dwork2014algorithmic}]\label{the:composition}
    Let $\mathtt{f}_i$ be a set of functions such that each provides $\epsilon_i$-differential privacy.
    First, the \emph{sequential composability} property of differential privacy states that computing all functions on the same dataset results in satisfying $(\sum_i \epsilon_i)$-differential privacy.
    Second, the \emph{parallel composability} property states that computing each function on disjoint subsets provides $\mathtt{max} ( \epsilon_i)$-differential privacy.
\end{theorem}

\section{Analysis of current practices}

In the following section, we review the current practice for legal data anonymization and privacy regulations.
We also make a connection with medical data anonymization techniques on which most papers rely.
To be concrete, we illustrate the privacy risks through examples of re-identification attacks.
Finally, we argue that rule-based anonymization is not sufficient to provide a strong privacy protection and discuss the (formal) issues surrounding text anonymization.

\subsection{Redaction \emph{in the wild}}\label{sec:redaction}

\paragraph{Redaction of legal data}

The redaction process consists in removing or generalizing a set of predefined terms defined by law through a semi-manual process~\cite{opijnen2017line}.
Furthermore, access to legal documents or even public hearings can be restricted in well-defined cases.
The common practice is to replace sensitive terms, as defined below, by initials, random letters, blanks or generalized terms (\eg, ``Montréal'' becomes ``Québec'').
The specific set of rules regarding protected terms and the associated replacement practice can differ between countries and courthouses~\cite{opijnen2017line}.

According to~\cite{Plamondon_04:Anonymisation}, the following information is to be systematically removed for any person (subject to a restriction on publication), as well as for each of his or her relatives (parents, children, teachers, neighbors, employers, colleagues, school \ldots):
\begin{enumerate}
    \item names,
    \item date and place of birth
    \item contact details (number, street, municipality, postal code, telephone, fax, email, web page, IP address),
    \item unique personal identifiers (social security number, health insurance number, medical file, passport, bank account, credit card, \ldots),
    \item personal possessions identifiers (license or serial number, cadastral designation, company name, \ldots)
\end{enumerate}
In some context, the following data is also removed if it can be used to identify one of the individuals aforementioned:
\begin{enumerate}\setcounter{enumi}{6}
    \item small communities or geographic locations,
    \item the accused and co-accused if their identity is not already protected by law,
    \item the intervenors (court experts, social workers, police officers, doctors, \ldots)
    \item unusual information (number of children if abnormally high, income if particularly high, exceptional occupation or function).
\end{enumerate}

\cite{conleySustainingPrivacyOpen} present numerous examples of legislation framing the publication of specific terms and putting restriction to the \emph{open-court principle}.
For instance, it is common by default to hide the identity of victims of sexual offenses or children in youth courts.
The identity of jurors and witnesses is also kept secret to avoid coercion or parties tailoring their strategy.
In addition in the USA, the fear for national security or the possible prejudice to another trial can lead to a complete ban on reporting being issued.

\paragraph{Paper versus digital}

The main difference between paper and digital access is the ``practical obscurity'' of paper records on the one hand, and the easy accessibility of digital records, on the other.
The awkwardness of accessing paper records stored in a public courthouse puts inherent limitations on the ability of individuals or groups to access those records.
In contrast, digital records are easy to analyze, can be searched in ``bulk'' by combining various key factors (\eg, divorce and children) and can potentially be accessed from any computer.
Thus, traditional distribution provides ``practical obscurity''~\cite{jtac2003open}, in that it is inconvenient (\ie, time-consuming) to attend the courthouse or read case reports.

\paragraph{Anonymization of medical data}

The Health Insurance Portability and Accountability Act (HIPAA) in the USA defines the security and privacy requirements of health information for both health professionals and technologies involved in medical data.
The search for complying with HIPPA has led to an important body of work on the redaction of health records.
In particular, automated redaction or generalization of the sensitive terms defined in HIPPA generally involves domain specific named-entity recognition and generalization of terms through medical ontologies.
As a concrete anonymization tool, Scrub~\cite{sweeney1996replacing} uses template matching to detect sensitive terms, which are replaced with synthetic data of similar type (\eg, a name with a name, a disease with a similar disease).
$t$-PAT~\cite{jiang2009tpat} replaces sensitive words or phrases--recognized by an ontology--with more general terms using an early privacy-preserving data publishing model, called $k$-anonymity~\cite{10.1142/S0218488502001648}, to preserve the privacy of patients.

\subsection{Limits of current approaches}\label{sec:limits}

Our objective in this section is to provide examples of potential attacks in order to illustrate the technical difficulties of raw text anonymization.
Figures~\ref{fig:Katopodis}, \ref{fig:initials}, \ref{fig:madrid}, \ref{fig:sango}, \ref{fig:youth} are excerpts from French and Canadian opinions\footnote{We translated them using DeepL (\url{https://www.deepl.com})}.

A common redaction practice is to replace names by initials as shown in Figure~\ref{fig:initials}.
The uniqueness of initials~\cite{finseth1993rfc} is increased by combining multiple parties, particularly if the relationship between the parties is known (\eg, in a divorce case).

\begin{figure}
    \begin{quoting}[font=itshape]
        \centering
        E.B. Petitioner v. V.I. Respondent \\[1ex]
        Judgment for Dissolution of Marriage
    \end{quoting}
    \caption{Droit de la famille -- 15334, 2015 QCCS 762 (CanLII), \url{http://canlii.ca/t/ggk9w}}\label{fig:initials}
\end{figure}

A combination of attributes, which can be extracted using dedicated named-entity recognition, is presented in Figure~\ref{fig:Katopodis}: names of parties, parties are divorced, date of marriage, parties have a daughter, birthdate of the daughter, date of divorce, reside in Ontario near a Dr. James. This combination could be used as a quasi-identifier by a re-identification attack.

\begin{figure}
    \begin{quoting}[font=itshape]
        {\centering
            Katopodis v. Katopodis \\[1ex]
            SUPREME COURT OF ONTARIO \\[1ex]}
        The parties were married on August 25, 1968; a daughter was born on November 30, 1972; the parties separated in April, 1977. The wife first went to see Dr. James
    \end{quoting}
    \caption{Katopodis v. Katopodis, 1979 CanLII 1887 (ON SC), \url{http://canlii.ca/t/g19bb}}\label{fig:Katopodis}
\end{figure}

Figure~\ref{fig:madrid} is anonymized according to the CNIL recommendations of 2006, which requires the last name of individuals to be replaced by its initial.
However, widely available background knowledge on the ``Real Madrid Club de Futbol'' combined with the (real-life) pseudonyms of the ``players'' trivially leaks their identity.

\begin{figure}
    \begin{quoting}[font=itshape]
        the association Real Madrid Club de Futbol and several players of this team, Zinedine Z., David B., Raul Gonzalès B. aka Raul, Ronaldo Luiz Nazario de L., aka Ronaldo, and Luis Filipe Madeira C., aka Luis Figo
    \end{quoting}
    \caption{CA Paris, 11\(^{\textnormal{e}}\) ch., sect. B, 14 February 2008, \emph{Unibet Ltd c/ Real Madrid et autres}, RG n\(^\circ\) 06/11504, GP}\label{fig:madrid}
\end{figure}

The de-anonymization of Figure~\ref{fig:sango} relies on the text semantics instead of background knowledge.
It requires the adversary (1) to identify the link (X) between ``M. [\dots] Abdel X'' and ``the use of the name `X' to designate a drink'', and (2) to infer that the drink is called ``sango'', thus leading to the conclusion that X $=$ ``sango''.
While this attack may not be easy to automatize due to the hardness of detecting the semantics inference, it is, however, trivial to perform for a human (\eg, by crowdsourcing it).

\begin{figure}
    \begin{quoting}[font=itshape]
        the American company Coca Cola Company markets drinks under the French trade mark "Coca Cola light sango", of which it is the proprietor; that M. [\dots] Abdel X, relying on the infringement of his artist's name and surname, has brought an action for damages against the Coca Cola Company [\dots] On the ground that Abdel X maintains that, as an author and screenwriter, he is entitled to oppose the use of the name ``X'' to designate a drink marketed by the companies of the Coca Cola group.
    \end{quoting}
    \caption{Civ. 1\(^{\textnormal{re}}\), 10 April 2013, n\(^\circ\) 12-14.525, \emph{Sango c/ Coca-Cola}, D. 2013. 992 ; CCE July 2013, n\(^\circ\) 73}\label{fig:sango}
\end{figure}

Similar to  Figure~\ref{fig:Katopodis}, Figure~\ref{fig:youth} could be attacked through a combination of attributes and relationship (\eg, extracted with Snorkel~\cite{ratner2020snorkel}).
This opinion from the Youth court involves children and, as such, follows the strictest anonymization rules of the SOQUIJ: only the year's birthdate of children is given and names are replaced by random letters.
However, an adversary can extract an extensive relationship graph (see Figure~\ref{fig:rel_graph_family}), which could be matched over a relationship database (\eg, Facebook). In this case, a quasi-identifier could be the relationship graph (or parts of it).

\begin{figure}
    \begin{quoting}[font=itshape]
        X, born [\dots] 2017; Y, born [\dots] 2018 the children and C; D the parents \\[1ex]
        Applications are submitted for X, aged 1 year, and Y, aged 2 months. The Director of Youth Protection (DYP) would like X to be entrusted to her aunt, Ms. E, until June 25, 2019. As for Y, that he be entrusted to a foster family for the next nine months. The father has two other children, Z and A, from his previous union with Mrs. F. The mother has another child, B, from her union with Mr. G.
    \end{quoting}
    \caption{Protection de la jeunesse -- 186470, 2018 QCCQ 6920 (SOQUIJ), \url{http://t.soquij.ca/x4L6N}}\label{fig:youth}
\end{figure}

\begin{figure}
    \centering
    \resizebox{\linewidth}{!}{
        \begin{tikzpicture}[
    font=\sffamily\scriptsize,
    node distance=1.5cm,
    ]
    \node (D) {D};
    \node[left=of D] (C) {C};
    \node[above=of D] (Y) {Y};
    \node[below=of C] (X) {X};
    \node[right=of D] (A) {A};
    \node[left=of C] (B) {B};
    \node[above=of C] (G) {G};
    \node[left=of X] (E) {E};
    \node[below=of A] (Z) {Z};
    \node[above=of A] (F) {F};
    \draw[->] (C) -- node[above, sloped]{mother} (Y);
    \draw[->] (C) -- node[above, sloped]{mother} (X);
    \draw[->] (C) -- node[above, sloped]{mother} (B);
    \draw[->] (D) -- node[above, sloped]{father} (Y);
    \draw[->] (D) -- node[above, sloped]{father} (X);
    \draw[->] (D) -- node[above, sloped]{father} (Z);
    \draw[->] (D) -- node[above, sloped]{father} (A);
    \draw[->] (F) edge[bend right=-40] node[above, sloped]{mother} (Z);
    \draw[->] (F) -- node[above, sloped]{mother} (A);
    \draw[->] (G) -- node[above, sloped]{father} (B);
    \draw[<->] (C) -- node[above, sloped]{sister} (E);
    \draw[<->] (C) -- node[above, sloped]{divorced} (G);
    \draw[<->] (D) -- node[above, sloped]{divorced} (F);
\end{tikzpicture}
    }
    \caption{Relationship graph manually extracted from Figure~\ref{fig:youth}}\label{fig:rel_graph_family}
\end{figure}
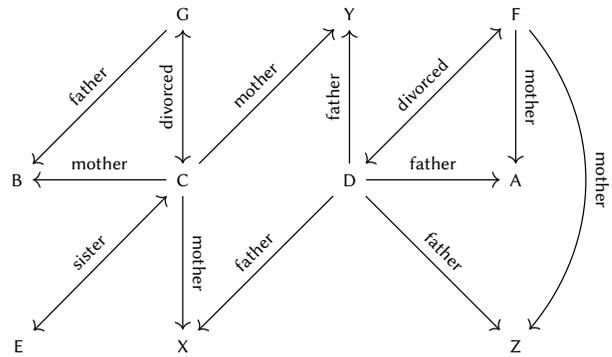

A study \cite{anandan2011significance} has shown that generalization-based sanitization is vulnerable to correlation attacks in which the generalized terms can be correlated with other terms, seemingly non-identifying, in order to jeopardize the effects of the generalization and consequently disclose sensitive terms.

Besides the content of legal documents, stylometry~\cite{neal2017surveying} can be used to identify authors (\ie, magistrates) by their writing style.
Mitigation for this kind of attack exist~\cite{fernandes2019generalised, weggenmann2018syntf} but their output is only machine readable.
Similarly, it is possible to exploit decision patterns to re-identify judges, as done for the Supreme Court of the United States~\cite{katz2017general}.

\subsection{Reasons for the failure of rule-based redaction}

\begin{figure*}[htbp]
    \centering
    \resizebox{\linewidth}{!}{
    \begin{tikzpicture}[
    >=latex,
    font=\sffamily\scriptsize,
    every node/.style={align=center},
    proc/.style={draw, fill=blue!15, rounded corners=2pt},
    obj/.style={fill=yellow!15},
    eg/.style={font=\itshape\scriptsize, text=black!60, inner sep=2pt, sloped, above},
    module/.style={draw, fill=green!15},
    ap/.style={font=\bfseries\sffamily\scriptsize},
    ]
    
    \node[fill=red!15, draw, double copy shadow] (a) {court records};
    \node[fit={(a.south west)([shift={(1ex,1ex)}]a.north east)}, inner sep=0] (db) {};
    
    \node[proc, above=of db] (mprep) {privacy-preserving preprocessing};
    \node[obj, right=of mprep] (mrepr) {privacy-preserving\\ structured representation};
    \draw[->] (db) -- (mprep)  -- (mrepr);
    
    \node[proc, below=of db] (redac) {redaction};
    \node[obj, right=of redac] (prepr) {redacted records};
    \draw[->] (db) -- (redac) -- (prepr);

    \node[left=3 of redac, module]  (privacy) {Privacy\\ parameters};
    \node[left=2 of mprep, module] (compose) {Composability\\ properties};
    \node[right=2.25 of db, module] (hist) {Disclosures log};

    \begin{scope}[on background layer]
        \draw[->] (privacy) -- node[eg] {max rate} (redac);
        \draw[->] (privacy) -- node[eg] {max $\epsilon$} (mprep);
        \draw[->] (hist) -- node[eg] {rate limiter} (redac);
        \draw[->] (hist) -- node[eg, below] {$\epsilon$ consumption} (mprep);
        \draw[->] (compose) -- node[eg] {sequential\\ and parallel} (mprep);
    \end{scope}
    
    \begin{scope}[on background layer]
        \node[draw, dashed, fit=(redac)(prepr), label={[ap]south:{precise AP}}] {};
        \node[draw, dashed, fit=(mprep)(mrepr), label={[ap]north:{massive AP}}] {};
    \end{scope}
    
    \node[bob, minimum size=0.75cm, right=2 of mrepr] (mu) {};
    \node[alice, minimum size=0.75cm] at (prepr -| mu) (pu) {};

    \coordinate (midu) at ($(mu)!0.50!(pu)$);
    
    \draw[->] (mrepr) -- (mu);
    \draw[->] (prepr) -- (pu);

    \draw[->, rounded corners=10pt] (mu) -- +(0,1.25) -| node[eg, pos=0.25, above] {preprocessing pipeline parameters} (mprep);

    \draw[->, rounded corners=10pt] (mu) -- +(0.75,-1.25) -- +(0.75,-4.2) -|  (privacy.-120);
    \draw[->, rounded corners=10pt] (pu) -- +(0,-1.05) -| (privacy.-60);
    
    \node[rotate=-90, anchor=north, minimum width=4.5cm, minimum height=1em] at ($(midu)+(-1.25,0)$) (ac) {};
    \fill[draw, fill=green!15, decoration={zigzag, segment length=3pt, amplitude=1pt}] decorate {(ac.south west) -- (ac.south east)} -- (ac.north east) decorate {-- (ac.north west)} -- cycle;
    \node[rotate=90] at (ac) (acl) {Authentication};
    
    \node at ($(mu.east)+(2.25,0)$) (out) {\dots};
    \draw[->] (mu.east) -- (out);
    \node[right=0.25 of mu, proc] (proc) {processing};
\end{tikzpicture}
    }
    \caption{Multimodal publication architecture}\label{fig:mult}
\end{figure*}
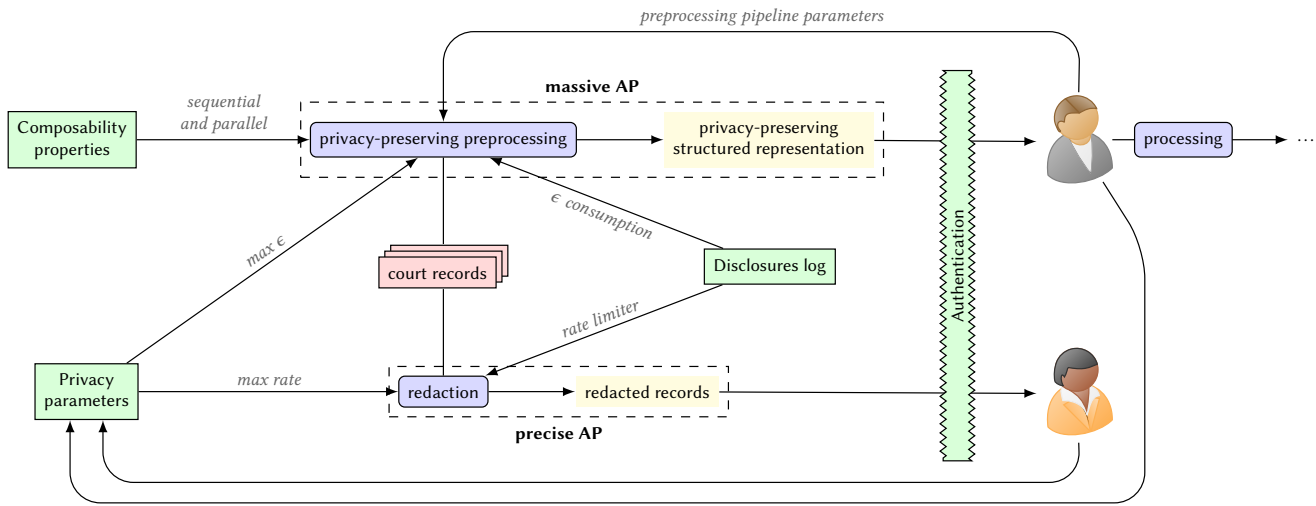

The review of current practices for tackling the privacy of legal documents in Section~\ref{sec:redaction} has highlighted the widespread use of rule-based redaction, in which a set of patterns is defined as being sensitive and is either removed or replaced.
However, as shown in Section~\ref{sec:limits} (1) privacy can be violated even in ``simple'' instances and (2) identifying information remains in most cases. In other words, rule-based redaction does not provide any sound privacy guarantee. 
We observe that it suffers from the following main difficulties.
\begin{enumerate}
    \item \emph{Missing rule difficulty}. Many combinations of quasi-identifiers can lead to re-identification and the richness of the output space offered by natural language (\ie, what can be expressed) can hardly be constrained to a set of rules.
    Furthermore, identifying the sensitive terms is challenging and domain-specific.
    This issue is the subject of multiple studies in the context of medical data~\cite{sweeney1996replacing, atallah2001natural, douglass2005identification}.
    \item \emph{Missing match difficulty}. The current state of the art about relationship extraction and named-entity recognition makes it hard to ensure that all terms that should be redacted will be detected, in particular because of the many possible ways to express the same idea (\eg, \emph{circumlocution}).
\end{enumerate}
Although these observations make the rule-based redaction difficult, it is important to note that attacks, \eg, re-identification, remain simpler than protection. Indeed, an adversary has to find a single attack vector (\ie, a missing rule or a missing pattern) whereas the redaction process needs to consider all the possibilities.

\section{Multimodal publication scheme}\label{sec:contrib}

In Section~\ref{sec:problem}, we have shown that the publication of legal documents serves two distinct and complementary purposes: (1) the traditional objective of transparency and case law, and (2) the modern objective of legal technologies of providing services to citizens and legal professionals.
These two purposes obey to different utility and privacy requirements.
More precisely, the traditional use case requires human-readable documents while legal techs need a machine-readable format for automated processing.
Moreover, transparency and case law involve the access to opinions on an individual basis (\ie, one-at-a-time), similarly to attending a hearing in person.
In contrast, legal technologies rely on the access to massive legal databases.
This difference in cardinality (\ie, one versus many) entails different privacy risks.
In particular, the massive processing of legal data requires the use of a formal privacy framework with composability properties (see Section~\ref{sec:privacy}). 
All this suggests the inadequacy of any \emph{one-size-fits-all} approach.

\subsection{Access modes}

As a consequence, we propose that the organization in charge of the publication of legal decisions should consider two modes of publication: the \emph{precise access mode} and the \emph{massive access mode}.

\paragraph{Precise access mode}

To fulfill the ``traditional'' use case, the precise access mode provides full access to legal decisions that are only redacted using the current practices.
This access mode is designed for the transparency and case law usages, and is to be used typically by individuals (\eg, law professionals, journalists and citizens).
Similar to the ``traditional'' paper-based publication scheme, in the precise access mode~\cite{hartzog2013case}, a user has access to text documents, either in full or only extracts (partial access is useful for crowdsourcing tagging in order to build a dataset).
While the current practice of redacting identifiers could be combined with more automated approaches such as~\cite{sanchez2013automatic, hassan2019automatic}. The aim of this mode is to provide strong utility first. It is similar to the websites currently publishing legal documents (\eg, Legifranceor CanLII), as it allows browsing, searching and reading documents.

To prevent malicious users from diverting the precise access mode for performing massive accesses, users must be authenticated and their access must be restricted (\eg, rate limitation or proof of work~\cite{dwork1992pricing}).
The main objective of the restricted access is to make it difficult to rebuild the full (massive) database. In addition, this mode provides privacy through ``practical obscurity'' similarly to the paper-based system.

\paragraph{Massive access mode}

The massive access mode gives access only to pre-processed data resulting from privacy-preserving versions of the standard NLP pipelines available on the server, \ie, 
 aggregated and structured data extracted from or computed over large numbers of decisions, as required for the ``modern'' use case. 
It should be compatible with most legal tech applications that traditionally use a database of legal documents (see Section~\ref{sec:legaltech}). Note that the perturbations due to privacy-preserving data publishing schemes have usually less impact (in terms of information loss) when applied late in the pipeline (see Figure~\ref{fig:mult}), at the cost of a loss of generality of the output.

Users need to be able to tune the pre-processing applied. For the sake of simplicity, we assume that the user (\ie, legaltech developer) provides the parameters for a given NLP pipeline (see Fig.~\ref{fig:mult}). These parameters can be for instance the maximum number of features or $n$-grams range to consider in the case of a BoW model or the window for word embeddings. But more complex implementations can be designed, \eg, allowing experiments by the users, fine-tuning for each dataset/task, as well as customization (\eg, for cleaning the data). This can be done (1) by generating structured synthetic \emph{testing} data (\eg, a set of features extracted from legal data) in a privacy-preserving manner (\eg, PATE-GAN~\cite{yoon2018pategan}) and (2) by designing a full pre-processing pipeline that embeds privacy-preserving calls to the server (\eg, through a privacy-preserving computation framework such as Ektelo~\cite{10.1145/3183713.3196921}). 

The massive access mode must also authenticate users in order to monitor the overall privacy guarantees satisfied for each user based on his disclosures log and on the composability properties of the privacy-preserving data publishing schemes used.

As a result, the data is protected using authentication and strong privacy definitions as presented in Section~\ref{sec:privacy}.
Examples of applications of differential privacy to NLP models include~\cite{fernandes2019generalised}, which adds noise to word-frequency-matrix to achieve differential privacy, or~\cite{weggenmann2018syntf}, which samples the dictionary of the model using the differentially-private exponential mechanism~\cite{dwork2014algorithmic}. The aim of these two approaches is to protect against authorship attribution.

\cite{fernandes2019generalised} uses a relaxation of differential privacy, $d_\chi$-privacy~\cite{alvim2018metric} which allows the authors to consider a distance between documents computed using word embeddings, rather than the row-based distance presented in~Definition~\ref{def:dp}.
The objective is to modify BoW representation of documents ``similar in topics'' remain ``similar to each other'' (w.r.t. the metrics defined on word embeddings), irrespective of authorship. 
In practice, this is achieved by drawing BoW where the probability of each word being associated to a document is distributed according to a Laplace probability density function.

The goal of~\cite{weggenmann2018syntf} is to derive a differentially private synthetic feature vectors, keeping the theme of each document while preventing authorship attribution. 
Feature vectors map a set of words (the dictionary) to probabilities of the word appearing in each document.
The main idea of the approach is to sample the dictionary from a reference dictionary (\eg, using synonyms from WordNet's synsets) using the differentially private exponential mechanism.

In practice, the massive access mode can be plugged into the existing platforms that store massive number of legal documents and already support the precise access mode, such as CourtListener or CanLII.

Finally, another potential need is the annotation of documents, which is the addition to terms, sentences, paragraphs or documents of metadata such as syntax information (\eg, verb or noun), semantic, pragmatic (\eg, presupposition and implicature).
This step is crucial in NLP, and is usually done manually, for example through crowdsourcing.
Crowdsourcing-specific approaches for privacy-preserving task processing~\cite{kajino2014instance} require to split the task (\ie, annotation of a set of documents) between non-colluding workers (\eg, at the sentence level) before aggregating the result.
Such approach is compatible with our architecture assuming the aggregation is done locally on the platform.

\subsection{System overview}

We now outline an abstract architecture for a privacy-preserving data publishing system for legal decisions.
Our objective is not to provide exhaustive implementation guidelines, but rather to identify the key components that such an architecture should possess.

Figure~\ref{fig:mult} depicts the proposed architecture.
The precise and massive access modes are both protected by the \texttt{Authentication} module.
The \texttt{Authentication} module can be implemented by usual strong authentication techniques (\eg, for preventing impersonation attacks).
Authentication is necessary for enforcing the access control policy through the \texttt{Access Control} module and for maintaining for each user his \texttt{Disclosure Log}.
The log contains all the successful access requests performed by a user.
It is required for verifying that the overall privacy guarantees are not breached, \eg, the rate limitation is not exceeded for the precise access mode, or the composition of the privacy-preserving data publishing schemes, formalized in the \texttt{Composability Properties}, does not exceed the disclosure allowed.
Finally, the \texttt{Privacy Parameters} contain the overall privacy guarantees that must always hold, defined by the administrator (\eg, rate limit or higher bound on the $\epsilon$ differential privacy parameter).
The user may additionally be allowed to tune the privacy parameters input by a privacy-preserving data publishing scheme (\eg, the fraction spent in the higher bound on the $\epsilon$ differential privacy parameter) provided it does not jeopardize the overall privacy guarantees.

\section{Discussion}\label{sec:conclusion}

In this paper, we analyzed the needs for publishing legal data and the limitations of rule-based redaction (\ie, the current approach) for fulfilling them successfully. We proposed to discard any one-size-fits-all approach and outlined a straw man architecture balancing the utility and privacy requirements by distinguishing the traditional, one-to-one, use of legal data from the modern, massive, use of legal data by legal technologies. Our proposition can easily be implemented on current platforms. 

\section{Acknowledgments}

This work was partially funded by the PROFILE-INT project funded by the LabEx CominLabs (ANR-10-LABX-07-01).
Sébastien Gambs is supported by the Canada Research Chair program as well as by a Discovery Grant from NSERC and the Legalia project from the AUDACE program funded by the FQRNT.

\bibliographystyle{plainnat}
\bibliography{main}

\begin{thebibliography}{72}
\providecommand{\natexlab}[1]{#1}
\providecommand{\url}[1]{\texttt{#1}}
\expandafter\ifx\csname urlstyle\endcsname\relax
  \providecommand{\doi}[1]{doi: #1}\else
  \providecommand{\doi}{doi: \begingroup \urlstyle{rm}\Url}\fi

\bibitem[Abbasi and Chen(2008)]{abbasi2008writeprints}
Ahmed Abbasi and Hsinchun Chen.
\newblock Writeprints: A stylometric approach to identity-level identification
  and similarity detection in cyberspace.
\newblock \emph{ACM Transactions on Information Systems (TOIS)}, 26\penalty0
  (2):\penalty0 7, 2008.
\newblock \doi{10.1145/1344411.1344413}.

\bibitem[Alarie et~al.(2018)Alarie, Niblett, and Yoon]{alarie2018artificial}
Benjamin Alarie, Anthony Niblett, and Albert~H Yoon.
\newblock How artificial intelligence will affect the practice of law.
\newblock \emph{University of Toronto Law Journal}, 68\penalty0 (supplement
  1):\penalty0 106--124, 2018.

\bibitem[Aletras et~al.(2016)Aletras, Tsarapatsanis, Preo{\c{t}}iuc-Pietro, and
  Lampos]{aletras2016predicting}
Nikolaos Aletras, Dimitrios Tsarapatsanis, Daniel Preo{\c{t}}iuc-Pietro, and
  Vasileios Lampos.
\newblock Predicting judicial decisions of the european court of human rights:
  A natural language processing perspective.
\newblock \emph{PeerJ Computer Science}, 2016.
\newblock \doi{10.7717/peerj-cs.93}.

\bibitem[Alvim et~al.(2018)Alvim, Chatzikokolakis, Palamidessi, and
  Pazii]{alvim2018metric}
M{\'a}rio~S Alvim, Konstantinos Chatzikokolakis, Catuscia Palamidessi, and Anna
  Pazii.
\newblock Metric-based local differential privacy for statistical applications.
\newblock \emph{arXiv preprint arXiv:1805.01456}, 2018.

\bibitem[Anandan and Clifton(2011)]{anandan2011significance}
Balamurugan Anandan and Chris Clifton.
\newblock Significance of term relationships on anonymization.
\newblock In \emph{Proceedings of the 2011 IEEE/WIC/ACM International
  Conferences on Web Intelligence and Intelligent Agent Technology-Volume 03},
  pages 253--256. IEEE Computer Society, 2011.

\bibitem[Arrington(2006)]{aol-arrington-tc-06}
Michael Arrington.
\newblock {AOL} {Proudly} {Releases} {Massive} {Amounts} of {Private} {Data}.
\newblock \emph{TechCrunch}, 2006.
\newblock URL \url{https://social.techcrunch.com/2006/08/06/aol-proudly-
  releases-massive-amounts-of-user-search-data/}.

\bibitem[Ashley and Br\"{u}ninghaus(2009)]{10.1007/s10506-009-9077-9}
Kevin~D. Ashley and Stefanie Br\"{u}ninghaus.
\newblock Automatically classifying case texts and predicting outcomes.
\newblock \emph{Artif. Intell. Law}, 17\penalty0 (2):\penalty0 125--165, June
  2009.
\newblock ISSN 0924-8463.
\newblock \doi{10.1007/s10506-009-9077-9}.

\bibitem[Ashley and Walker(2013)]{ashley2013toward}
Kevin~D Ashley and Vern~R Walker.
\newblock Toward constructing evidence-based legal arguments using legal
  decision documents and machine learning.
\newblock In \emph{Proceedings of the Fourteenth International Conference on
  Artificial Intelligence and Law}, pages 176--180, 2013.

\bibitem[Atallah et~al.(2001)Atallah, McDonough, Raskin, and
  Nirenburg]{atallah2001natural}
Mikhail~J Atallah, Craig~J McDonough, Victor Raskin, and Sergei Nirenburg.
\newblock Natural language processing for information assurance and security:
  an overview and implementations.
\newblock In \emph{Proceedings of the 2000 workshop on New security paradigms},
  pages 51--65, 2001.

\bibitem[Bailey and Burkell(2016)]{bailey2016revisiting}
Jane Bailey and Jacquelyn Burkell.
\newblock Revisiting the open court principle in an era of online publication:
  Questioning presumptive public access to parties' and witnesses' personal
  information.
\newblock \emph{Ottawa L. Rev.}, 48:\penalty0 143, 2016.

\bibitem[Bentham and Bowring(1843)]{bentham1843works}
Jeremy Bentham and John Bowring.
\newblock \emph{The Works of Jeremy Bentham}, volume~4.
\newblock W. Tait, 1843.

\bibitem[Br{\"u}ninghaus and Ashley(1997)]{10.1007/3-540-63233-6_501}
Stefanie Br{\"u}ninghaus and Kevin~D. Ashley.
\newblock Using machine learning for assigning indices to textual cases.
\newblock In David~B. Leake and Enric Plaza, editors, \emph{Case-Based
  Reasoning Research and Development}, pages 303--314, Berlin, Heidelberg,
  1997. Springer Berlin Heidelberg.
\newblock ISBN 978-3-540-69238-6.

\bibitem[Calamur(2014)]{secret-trial-britain}
Krishnadev Calamur.
\newblock In a first for {Britain}, a secret trial for terrorism suspects.
\newblock \emph{NPR}, 2014.
\newblock URL \url{https://text.npr.org/s.php?sId=319076959}.

\bibitem[Chen et~al.(2009)Chen, Kifer, LeFevre, and
  Machanavajjhala]{ppdp-chen-ft-09}
Bee-Chung Chen, Daniel Kifer, Kristen LeFevre, and Ashwin Machanavajjhala.
\newblock Privacy-preserving data publishing.
\newblock \emph{Found. Trends Databases}, 2\penalty0 (1--2):\penalty0 1--167,
  January 2009.
\newblock ISSN 1931-7883.
\newblock \doi{10.1561/1900000008}.

\bibitem[Committee(2003)]{jtac2003open}
Judges Technology~Advisory Committee.
\newblock Open courts, electronic access to court records, and privacy:
  discussion paper.
\newblock Technical report, Canadian Judicial Council, 2003.
\newblock URL \url{http://publications.gc.ca/collections/collection_2008/lcc-
  cdc/JL2-75-2003E.pdf}.

\bibitem[Conley et~al.(2011)Conley, Datta, Nissenbaum, and
  Sharma]{conleySustainingPrivacyOpen}
Amanda Conley, Anupam Datta, Helen Nissenbaum, and Divya Sharma.
\newblock Sustaining privacy and open justice in the transition to online court
  records: A multidisciplinary inquiry.
\newblock \emph{Md. L. Rev.}, 71:\penalty0 772, 2011.

\bibitem[Custis et~al.(2019)Custis, Schilder, Vacek, McElvain, and
  Alonso]{custisWestlawEdgeAI2019}
Tonya Custis, Frank Schilder, Thomas Vacek, Gayle McElvain, and Hector~Martinez
  Alonso.
\newblock Westlaw {{Edge AI Features Demo}}: {{KeyCite Overruling Risk}},
  {{Litigation Analytics}}, and {{WestSearch Plus}}.
\newblock In \emph{Proceedings of the {{Seventeenth International Conference}}
  on {{Artificial Intelligence}} and {{Law}} - {{ICAIL}} '19}, pages 256--257,
  Montreal, QC, Canada, 2019. ACM Press.
\newblock ISBN 978-1-4503-6754-7.
\newblock \doi{10.1145/3322640.3326739}.
\newblock URL \url{http://dl.acm.org/citation.cfm?doid=3322640.3326739}.

\bibitem[Dale(2019)]{dale_law_2019}
Robert Dale.
\newblock Law and word order: {NLP} in legal tech.
\newblock \emph{Natural Language Engineering}, 25\penalty0 (1):\penalty0
  211--217, January 2019.
\newblock \doi{10.1017/s1351324918000475}.
\newblock URL \url{https://www.cambridge.org/core/product/identifier/
  S1351324918000475/type/journal%5Farticle}.

\bibitem[Declaration on Free Access to Law()]{fal}
Declaration on Free Access to Law.
\newblock Declaration on free access to law, 2002.
\newblock URL \url{http://www.worldlii.org/worldlii/declaration/}.

\bibitem[Douglass et~al.(2005)Douglass, Cliffford, Reisner, Long, Moody, and
  Mark]{douglass2005identification}
MM~Douglass, GD~Cliffford, Andrew Reisner, WJ~Long, GB~Moody, and RG~Mark.
\newblock De-identification algorithm for free-text nursing notes.
\newblock In \emph{Computers in Cardiology, 2005}, pages 331--334. IEEE, 2005.

\bibitem[Dwork(2006)]{Dwork:2006:DP:2097282.2097284}
Cynthia Dwork.
\newblock Differential privacy.
\newblock In \emph{Proceedings of the 33\textsuperscript{rd} International
  Conference on Automata, Languages and Programming - Volume Part II}, volume
  4052 of \emph{Icalp'06}, pages 1--12, Berlin, Heidelberg, July 2006.
  Springer-Verlag.
\newblock ISBN 3-540-35907-9, 978-3-540-35907-4.
\newblock \doi{10.1007/11787006\_1}.
\newblock URL \url{https://www.microsoft.com/en-us/research/publication/
  differential-privacy/}.

\bibitem[Dwork and Naor(1992)]{dwork1992pricing}
Cynthia Dwork and Moni Naor.
\newblock Pricing via processing or combatting junk mail.
\newblock In \emph{Annual International Cryptology Conference}, pages 139--147.
  Springer, 1992.

\bibitem[Dwork et~al.(2014)Dwork, Roth, et~al.]{dwork2014algorithmic}
Cynthia Dwork, Aaron Roth, et~al.
\newblock The algorithmic foundations of differential privacy.
\newblock \emph{Foundations and Trends{\textregistered} in Theoretical Computer
  Science}, 9\penalty0 (3--4):\penalty0 211--407, 2014.

\bibitem[Fernandes et~al.(2019)Fernandes, Dras, and
  McIver]{fernandes2019generalised}
Natasha Fernandes, Mark Dras, and Annabelle McIver.
\newblock Generalised differential privacy for text document processing.
\newblock In \emph{International Conference on Principles of Security and
  Trust}, pages 123--148. Springer, 2019.
\newblock \doi{10.1007/978-3-030-17138-4\_6}.

\bibitem[Finseth(1993)]{finseth1993rfc}
Craig Finseth.
\newblock Rfc 1439 uniqueness of unique identifiers.
\newblock RFC 1439, {RFC Editor}, March 1993.

\bibitem[Fleuriot(2017)]{fleuriot2017avec}
Caroline Fleuriot.
\newblock Avec l'acc{\`e}s gratuit {\`a} toute la jurisprudence, des magistrats
  r{\'e}clament l'anonymat.
\newblock \emph{Dalloz Actualit{\'e}}, February 2017.
\newblock URL \url{https://www.dalloz-actualite.fr/flash/avec-l-acces-gratuit-
  toute-jurisprudence-des-magistrats-reclament-l-anonymat}.

\bibitem[for Economic Co-operation and
  Development(2011)]{organisation_for_economic_co-operation_and_development_call_2011}
Organisation for Economic Co-operation and Development, editors.
\newblock \emph{The call for innovative and open government: an overview of
  country initiatives}.
\newblock Oecd, Paris, 2011.
\newblock ISBN 9789264107045.

\bibitem[Fouret et~al.(2019)Fouret, Perez, Barri{\`{e}}re, Rottier, and
  Buat-M{\'{e}}nard]{etalab-openjustice}
Amaury Fouret, Mathieu Perez, Valentin Barri{\`{e}}re, Edouard Rottier, and
  {{\'{E}}}loi Buat-M{\'{e}}nard.
\newblock {Open} {Justice}.
\newblock Technical report, Cour de cassation, 2019.
\newblock URL \url{https://entrepreneur-interet-general.etalab.gouv.fr/defis/
  2019/openjustice.html}.

\bibitem[Hartzog and Stutzman(2013)]{hartzog2013case}
Woodrow Hartzog and Frederic Stutzman.
\newblock The case for online obscurity.
\newblock \emph{Calif. L. Rev.}, 101:\penalty0 1, 2013.

\bibitem[Hassan et~al.(2019)Hassan, S\'{a}nchez, Soria-Comas, and
  Domingo-Ferrer]{hassan2019automatic}
Fadi Hassan, David S\'{a}nchez, Jordi Soria-Comas, and Josep Domingo-Ferrer.
\newblock Automatic {{Anonymization}} of {{Textual Documents}}: {{Detecting
  Sensitive Information}} via {{Word Embeddings}}.
\newblock In \emph{2019 18\textsuperscript{th} {{IEEE International Conference
  On Trust}}, {{Security And Privacy In Computing And
  Communications}}/13\textsuperscript{th} {{IEEE International Conference On
  Big Data Science And Engineering}} ({{TrustCom}}/{{BigDataSE}})}, pages
  358--365, 2019.
\newblock \doi{10.1109/TrustCom/BigDataSE.2019.00055}.

\bibitem[Jaconelli(2002)]{jaconelli2002open}
Joseph Jaconelli.
\newblock \emph{Open Justice: A critique of the public trial}.
\newblock Oxford University Press on Demand, 2002.

\bibitem[Jacquin(2017)]{jacquin2017peur}
Jean-Baptiste Jacquin.
\newblock Terrorisme~: la peur des magistrats.
\newblock \emph{Le Monde}, January 2017.
\newblock URL \url{https://www.lemonde.fr/police-justice/article/2017/01/19/
  terrorisme-la-peur-des-magistrats_5065242_1653578.html}.

\bibitem[Jiang et~al.(2009)Jiang, Murugesan, Clifton, and Si]{jiang2009tpat}
Wei Jiang, Mummoorthy Murugesan, Chris Clifton, and Luo Si.
\newblock t-plausibility: Semantic preserving text sanitization.
\newblock In \emph{2009 International Conference on Computational Science and
  Engineering}, volume~3, pages 68--75. IEEE, 2009.

\bibitem[Joachims(1998)]{joachims1998text}
Thorsten Joachims.
\newblock Text categorization with support vector machines: Learning with many
  relevant features.
\newblock In \emph{European conference on machine learning}, pages 137--142.
  Springer, 1998.
\newblock \doi{10.1007/bfb0026683}.

\bibitem[Jordon et~al.(2019)Jordon, Yoon, and van~der Schaar]{yoon2018pategan}
James Jordon, Jinsung Yoon, and Mihaela van~der Schaar.
\newblock {PATE-GAN:} generating synthetic data with differential privacy
  guarantees.
\newblock In \emph{7\textsuperscript{th} International Conference on Learning
  Representations, {ICLR} 2019, New Orleans, LA, USA, May 6-9, 2019}.
  OpenReview.net, 2019.
\newblock URL \url{https://openreview.net/forum?id=S1zk9iRqF7}.

\bibitem[Joulin et~al.(2016)Joulin, Grave, Bojanowski, and
  Mikolov]{joulin2016bag}
Armand Joulin, Edouard Grave, Piotr Bojanowski, and Tomas Mikolov.
\newblock Bag of tricks for efficient text classification.
\newblock \emph{arXiv preprint arXiv:1607.01759}, 2016.
\newblock \doi{10.18653/v1/e17-2068}.

\bibitem[Kajino et~al.(2014)Kajino, Baba, and Kashima]{kajino2014instance}
Hiroshi Kajino, Yukino Baba, and Hisashi Kashima.
\newblock Instance-privacy preserving crowdsourcing.
\newblock In \emph{Second AAAI Conference on Human Computation and
  Crowdsourcing}, 2014.

\bibitem[Katz et~al.(2017)Katz, Bommarito~II, and Blackman]{katz2017general}
Daniel~Martin Katz, Michael~J. Bommarito~II, and Josh Blackman.
\newblock A general approach for predicting the behavior of the supreme court
  of the united states.
\newblock \emph{PLoS One}, 12\penalty0 (4), 2017.
\newblock \doi{10.1371/journal.pone.0174698}.

\bibitem[Kim et~al.(2019)Kim, Rabelo, and Goebel]{kimStatuteLawInformation2019}
Mi-Young Kim, Juliano Rabelo, and Randy Goebel.
\newblock Statute {{Law Information Retrieval}} and {{Entailment}}.
\newblock In \emph{Proceedings of the {{Seventeenth International Conference}}
  on {{Artificial Intelligence}} and {{Law}} - {{ICAIL}} '19}, pages 283--289,
  Montreal, QC, Canada, 2019. ACM Press.
\newblock ISBN 978-1-4503-6754-7.
\newblock \doi{10.1145/3322640.3326742}.
\newblock URL \url{http://dl.acm.org/citation.cfm?doid=3322640.3326742}.

\bibitem[Kim(2014)]{kim2014convolutional}
Yoon Kim.
\newblock Convolutional neural networks for sentence classification.
\newblock \emph{arXiv preprint arXiv:1408.5882}, 2014.
\newblock \doi{10.3115/v1/d14-1181}.

\bibitem[Kort(1965)]{kort1965quantitative}
Fred Kort.
\newblock Quantitative analysis of fact-patterns in cases and their impact on
  judicial decisions.
\newblock \emph{Harv. L. Rev.}, 79:\penalty0 1595, 1965.

\bibitem[Lai et~al.(2015)Lai, Xu, Liu, and Zhao]{lai2015recurrent}
Siwei Lai, Liheng Xu, Kang Liu, and Jun Zhao.
\newblock Recurrent convolutional neural networks for text classification.
\newblock In \emph{Twenty-ninth AAAI conference on artificial intelligence},
  2015.

\bibitem[Liu et~al.(2016)Liu, Qiu, and Huang]{liu2016recurrent}
Pengfei Liu, Xipeng Qiu, and Xuanjing Huang.
\newblock Recurrent neural network for text classification with multi-task
  learning.
\newblock \emph{arXiv preprint arXiv:1605.05101}, 2016.

\bibitem[Mandal et~al.(2017)Mandal, Chaki, Saha, Ghosh, Pal, and
  Ghosh]{10.1145/3140107.3140119}
Arpan Mandal, Raktim Chaki, Sarbajit Saha, Kripabandhu Ghosh, Arindam Pal, and
  Saptarshi Ghosh.
\newblock Measuring similarity among legal court case documents.
\newblock In \emph{Proceedings of the 10\textsuperscript{th} Annual ACM India
  Compute Conference}, Compute {\textquoteright}17, pages 1--9, New York, NY,
  USA, 2017. Association for Computing Machinery.
\newblock ISBN 9781450353236.
\newblock \doi{10.1145/3140107.3140119}.

\bibitem[Marques et~al.(2019)Marques, Bianco, Roodnejad, Baduel, and
  Berrou]{marquesMachineLearningExplaining2019}
Max~RS Marques, Tommaso Bianco, Maxime Roodnejad, Thomas Baduel, and Claude
  Berrou.
\newblock Machine learning for explaining and ranking the most influential
  matters of law.
\newblock In \emph{Proceedings of the Seventeenth International Conference on
  Artificial Intelligence and Law}, pages 239--243. Acm, 2019.

\bibitem[Marrero et~al.(2013)Marrero, Urbano, S{\'a}nchez-Cuadrado, Morato, and
  G{\'o}mez-Berb{\'\i}s]{marrero2013named}
M{\'o}nica Marrero, Juli{\'a}n Urbano, Sonia S{\'a}nchez-Cuadrado, Jorge
  Morato, and Juan~Miguel G{\'o}mez-Berb{\'\i}s.
\newblock Named entity recognition: fallacies, challenges and opportunities.
\newblock \emph{Computer Standards \& Interfaces}, 35\penalty0 (5):\penalty0
  482--489, 2013.

\bibitem[Martin(2008)]{martin2008online}
Peter~W Martin.
\newblock Online access to court records-from documents to data, particulars to
  patterns.
\newblock \emph{Vill. L. Rev.}, 53:\penalty0 855, 2008.

\bibitem[McClean(2011)]{mcclean2011not}
Tom McClean.
\newblock Not with a bang but a whimper: The politics of accountability and
  open data in the uk.
\newblock In \emph{APSA 2011 Annual Meeting Paper}, 2011.

\bibitem[McDermott(2010)]{mcdermott2010building}
Patrice McDermott.
\newblock Building open government.
\newblock \emph{Government Information Quarterly}, 27\penalty0 (4):\penalty0
  401--413, 2010.

\bibitem[Mikolov et~al.(2013)Mikolov, Sutskever, Chen, Corrado, and
  Dean]{mikolov2013distributed}
Tomas Mikolov, Ilya Sutskever, Kai Chen, Greg~S Corrado, and Jeff Dean.
\newblock Distributed representations of words and phrases and their
  compositionality.
\newblock In \emph{Advances in neural information processing systems}, pages
  3111--3119, 2013.

\bibitem[Minocha and Singh(2018)]{MINOCHA189}
Akshay Minocha and Navjyoti Singh.
\newblock Legal document similarity using triples extracted from unstructured
  text.
\newblock In Georg Rehm, V{\'{i}}ctor Rodr{\'{i}}guez-Doncel, and Juli{\'{a}}n
  Moreno-Schneider, editors, \emph{Proceedings of the Eleventh International
  Conference on Language Resources and Evaluation (LREC 2018)}, Paris, France,
  May 2018. European Language Resources Association (ELRA).
\newblock ISBN 979-10-95546-18-4.

\bibitem[Mokanov et~al.(2019)Mokanov, Shane, and Cerat]{mokanov2019facts2law}
Ivan Mokanov, Daniel Shane, and Benjamin Cerat.
\newblock Facts2law: using deep learning to provide a legal qualification to a
  set of facts.
\newblock In \emph{Proceedings of the Seventeenth International Conference on
  Artificial Intelligence and Law}, pages 268--269. Acm, 2019.

\bibitem[Mo{\v{z}}ina et~al.(2005)Mo{\v{z}}ina, {\v{Z}}abkar, Bench-Capon, and
  Bratko]{movzina2005argument}
Martin Mo{\v{z}}ina, Jure {\v{Z}}abkar, Trevor Bench-Capon, and Ivan Bratko.
\newblock Argument based machine learning applied to law.
\newblock \emph{Artificial Intelligence and Law}, 13\penalty0 (1):\penalty0
  53--73, 2005.

\bibitem[Nallapati and Manning(2008)]{nallapati2008legal}
Ramesh Nallapati and Christopher~D Manning.
\newblock Legal docket classification: Where machine learning stumbles.
\newblock In \emph{Proceedings of the 2008 Conference on Empirical Methods in
  Natural Language Processing}, pages 438--446, 2008.

\bibitem[Neal et~al.(2017)Neal, Sundararajan, Fatima, Yan, Xiang, and
  Woodard]{neal2017surveying}
Tempestt Neal, Kalaivani Sundararajan, Aneez Fatima, Yiming Yan, Yingfei Xiang,
  and Damon Woodard.
\newblock Surveying stylometry techniques and applications.
\newblock \emph{ACM Computing Surveys (CSUR)}, 50\penalty0 (6):\penalty0 1--36,
  2017.

\bibitem[of~{Houston} {Law}~{Center}(2009)]{brief-case}
{University} of~{Houston} {Law}~{Center}.
\newblock How to brief a case.
\newblock Technical report, {University} of {Houston} {Law} {Center}, 2009.
\newblock URL \url{https://www.law.uh.edu/lss/casebrief.pdf}.

\bibitem[Opijnen et~al.(2017)Opijnen, Peruginelli, Kefali, and
  Palmirani]{opijnen2017line}
Marc Opijnen, Ginevra Peruginelli, Eleni Kefali, and Monica Palmirani.
\newblock On-line publication of court decisions in the {EU}: Report of the
  policy group of the project ``building on the european case law identifier''.
\newblock \emph{Available at SSRN 3088495}, 2017.

\bibitem[Plamondon et~al.(2004)Plamondon, Lapalme, and
  Pelletier]{Plamondon_04:Anonymisation}
Luc Plamondon, Guy Lapalme, and Fr{\'e}d{\'e}ric Pelletier.
\newblock Anonymisation de d{\'e}cisions de justice.
\newblock In \emph{XIe Conf{\'e}rence sur le Traitement Automatique des Langues
  Naturelles (TALN 2004).}, pages 367--376, F{\`e}s, Maroc, May 2004. Bernard
  Bel et Isabelle Martin. ({\'e}diteurs), Bernard Bel et Isabelle Martin.
  ({\'e}diteurs).
\newblock URL \url{http://rali.iro.umontreal.ca/rali/sites/default/files/
  publis/0UdeM-taln-04.pdf}.

\bibitem[Praduroux et~al.(2016)Praduroux, de~Paiva, and
  di~Caro]{praduroux2016legal}
Sabrina Praduroux, Valeria de~Paiva, and Luigi di~Caro.
\newblock Legal tech start-ups: State of the art and trends.
\newblock In \emph{Proceedings of the Workshop on MIning and REasoning with
  Legal texts collocated at the 29\textsuperscript{th} International Conference
  on Legal Knowledge and Information Systems}, 2016.

\bibitem[Quaresma and Gon{\c{c}}alves(2010)]{Quaresma2010}
Paulo Quaresma and Teresa Gon{\c{c}}alves.
\newblock Using linguistic information and machine learning techniques to
  identify entities from juridical documents.
\newblock In Enrico Francesconi, Simonetta Montemagni, Wim Peters, and Daniela
  Tiscornia, editors, \emph{Semantic Processing of Legal Texts: Where the
  Language of Law Meets the Law of Language}, pages 44--59. Springer Berlin
  Heidelberg, Berlin, Heidelberg, 2010.
\newblock ISBN 978-3-642-12837-0.
\newblock \doi{10.1007/978-3-642-12837-0_3}.

\bibitem[Ratner et~al.(2020)Ratner, Bach, Ehrenberg, Fries, Wu, and
  R{\'e}]{ratner2020snorkel}
Alexander Ratner, Stephen~H Bach, Henry Ehrenberg, Jason Fries, Sen Wu, and
  Christopher R{\'e}.
\newblock Snorkel: Rapid training data creation with weak supervision.
\newblock \emph{The VLDB Journal}, 29\penalty0 (2):\penalty0 709--730, 2020.

\bibitem[Sanchez et~al.(2013)Sanchez, Batet, and Viejo]{sanchez2013automatic}
David Sanchez, Montserrat Batet, and Alexandre Viejo.
\newblock Automatic general-purpose sanitization of textual documents.
\newblock \emph{IEEE Transactions on Information Forensics and Security},
  8\penalty0 (6):\penalty0 853--862, 2013.
\newblock \doi{10.1109/tifs.2013.2239641}.

\bibitem[Siegel(2017)]{siegel2017cara}
Daniel~J Siegel.
\newblock Cara: An assistance to help find the cases you missed.
\newblock \emph{Law Prac.}, 43:\penalty0 22, 2017.

\bibitem[Sweeney(1996)]{sweeney1996replacing}
Latanya Sweeney.
\newblock Replacing personally-identifying information in medical records, the
  scrub system.
\newblock In \emph{Proceedings of the AMIA annual fall symposium}, page 333.
  American Medical Informatics Association, 1996.

\bibitem[Sweeney(2002)]{10.1142/S0218488502001648}
Latanya Sweeney.
\newblock K-anonymity: A model for protecting privacy.
\newblock \emph{Int. J. Uncertain. Fuzziness Knowl.-Based Syst.}, 10\penalty0
  (5):\penalty0 557--570, October 2002.
\newblock ISSN 0218-4885.
\newblock \doi{10.1142/S0218488502001648}.

\bibitem[Tan et~al.(2018)Tan, Adebayo, Inkpen, and Kamar]{tan2018investigating}
Sarah Tan, Julius Adebayo, Kori Inkpen, and Ece Kamar.
\newblock Investigating human+ machine complementarity for recidivism
  predictions.
\newblock \emph{arXiv preprint arXiv:1808.09123}, 2018.

\bibitem[Thenmozhi et~al.(2017)Thenmozhi, Kannan, and
  Aravindan]{thenmozhi2017text}
D~Thenmozhi, Kawshik Kannan, and Chandrabose Aravindan.
\newblock A text similarity approach for precedence retrieval from legal
  documents.
\newblock In \emph{FIRE (Working Notes)}, pages 90--91, 2017.

\bibitem[Weggenmann and Kerschbaum(2018)]{weggenmann2018syntf}
Benjamin Weggenmann and Florian Kerschbaum.
\newblock Syntf: Synthetic and differentially private term frequency vectors
  for privacy-preserving text mining.
\newblock \emph{arXiv preprint arXiv:1805.00904}, 2018.
\newblock \doi{10.1145/3209978.3210008}.

\bibitem[Yousfi-Monod et~al.(2010)Yousfi-Monod, Farzindar, and
  Lapalme]{10.1007/978-3-642-13059-5_8}
Mehdi Yousfi-Monod, Atefeh Farzindar, and Guy Lapalme.
\newblock Supervised machine learning for summarizing legal documents.
\newblock In Atefeh Farzindar and Vlado Ke{\v{s}}elj, editors, \emph{Advances
  in Artificial Intelligence}, pages 51--62, Berlin, Heidelberg, 2010. Springer
  Berlin Heidelberg.
\newblock ISBN 978-3-642-13059-5.

\bibitem[Zhang et~al.(2018)Zhang, McKenna, Kotsogiannis, Hay, Machanavajjhala,
  and Miklau]{10.1145/3183713.3196921}
Dan Zhang, Ryan McKenna, Ios Kotsogiannis, Michael Hay, Ashwin Machanavajjhala,
  and Gerome Miklau.
\newblock Ektelo: A framework for defining differentially-private computations.
\newblock In \emph{Proceedings of the 2018 International Conference on
  Management of Data}, SIGMOD {\textquoteright}18, pages 115--130, New York,
  NY, USA, 2018. Association for Computing Machinery.
\newblock ISBN 9781450347037.
\newblock \doi{10.1145/3183713.3196921}.

\bibitem[Zhang et~al.(2015)Zhang, Zhao, and LeCun]{zhang2015character}
Xiang Zhang, Junbo Zhao, and Yann LeCun.
\newblock Character-level convolutional networks for text classification.
\newblock In \emph{Advances in neural information processing systems}, pages
  649--657, 2015.

\bibitem[Zheng et~al.(2018)Zheng, Guo, Feng, and Chen]{zheng2018hierarchical}
Jianming Zheng, Yupu Guo, Chong Feng, and Honghui Chen.
\newblock A hierarchical neural-network-based document representation approach
  for text classification.
\newblock \emph{Mathematical Problems in Engineering}, 2018, 2018.
\newblock \doi{10.1155/2018/7987691}.

\end{thebibliography}

\end{document}